\documentclass[aps,pra,nofootinbib,preprint,amsmath,amssymb,floatfix]{revtex4-2}
\usepackage{color}
\usepackage{graphicx}

\begin{document}

\newcommand{\tc}{\textcolor}
\newcommand{\g}{blue}
\newcommand{\ve}{\varepsilon}
\title{Fluctuational Electrodynamics in and out of Equilibrium  }         
\author{ Iver Brevik$^1$  }      
\affiliation{$^1$Department of Energy and Process Engineering, Norwegian University of Science and Technology, N-7491 Trondheim, Norway}
\date{\today}          
\author{Boris Shapiro$^2$ }
\affiliation{$^2$  Department of Physics, Technion, Israel Institute of Technology, Haifa, Israel }

\author{M\'ario G. Silveirinha$^3$}
\affiliation{$^3$ University of Lisbon and Instituto de Telecomunicações, Avenida Rovisco Pais 1, Lisboa, 1049-001 Portugal}	

\begin{abstract}
   Dispersion  forces between neutral material bodies  are due to fluctuations of the polarization of the bodies. For bodies in equilibrium these forces are often referred  to as Casimir-Lifshitz forces. For bodies in relative motion, in addition to the Casimir-Lifshitz force, a lateral  frictional force ("quantum friction", in the zero temperature limit) comes into play. The widely accepted theory of the fluctuation induced forces is based on the "fluctuational electrodynamics" , when the Maxwell equations are supplemented by random current sources responsible for the fluctuations of the medium polarization.

  The first part of our paper touches on some conceptual issues of the theory, such as the dissipation-less limit and the link between Rytov's approach and quantum electrodynamics. We point out the problems with  the dissipation-less plasma model (with its unphysical double pole at zero frequency) which still appears in the literature.
    The second part of the paper is devoted to "quantum friction", in a broad sense, and it contains some novel material. In particular, it is pointed out that in  weakly dissipative systems the friction force may not be a stationary process. It is shown, using an "exact" (nonpertubative) quantum treatment that under appropriate conditions, an instability can occur when the kinetic energy (due to the relative motion between the bodies) is transformed into coherent radiation, exponentially growing in intensity (the instability gets eventually limited by non-linear effects). We also discuss a setup when the two bodies are at rest but a constant electric current is flowing in one of the bodies. One may say that only the electron component of one body  is dragged with respect to the other body, unlike the usual setup when the two bodies are in relative motion. Clearly there are differences in the frictional forces between the two setups.
\end{abstract}
\maketitle

\bigskip
\section{Introduction}
\label{secintro}

\text Material bodies, in thermal equilibrium with the environment at some temperature $T$, maintain in their interior fluctuating currents ${{\bf j}^{\rm{f}}({{\bf r}}, t)}$. The fluctuations have both quantum and thermal origin, and at $T=0$ only the quantum fluctuations survive (zero-point motion). The fluctuation-dissipation theorem (FDT) allows for a quantitative description of these currents, by expressing the correlation function $\langle {{\bf j}^{\rm f}_i({{\bf r}}, t)}{{\bf j}^{\rm f}_k({{\bf r'}}, t')}\rangle$ in terms of the dielectric function of the body and the temperature. The indices $i, k$ above label the components of the current density and the angular brackets indicate the quantum and thermal average.

The fluctuational electrodynamics, initiated by Rytov  \cite{rytov89,Levin67,landau60,intra,GB84,kardar,volo,dedkov} amounts to augmenting the standard Maxwell equations in the medium by the fluctuating currents, as sources. One can then compute the correlation functions among various components of the electric and magnetic fields, produced by those fluctuation currents  and, thus (with the help of the Maxwell stress tensor), the forces acting on any of the bodies.
A textbook example of such force is due to Lifshitz \cite{lifshitz} who considered a simple geometry of two half-spaces, with dielectric constants $\epsilon_1$ and $\epsilon_2$ respectively, separated by a vacuum gap. The resulting expression is often called ``the Lifshitz formula'', which in the appropriate limit  yields the famous Casimir result \cite{casimir48}. More generally, one can consider $N$ material bodies and inquire about the (Casimir-Lifshitz) forces among the bodies. The correlation functions of the field components are determined by the correlator of the fluctuating currents which, in the frequency domain, is
\begin{align}
 \langle {{\bf j}^{\rm f}_i({\bf r}, \omega)}{{\bf j}^{\rm f}_k({\bf r'}, \omega')^*}\rangle  = \frac{1}{2}\hbar \omega^2\coth \left( \frac{\hbar \omega}{2T}\right){\rm{Im}} \, \varepsilon({\rm r, \omega})\delta_{ik}\delta({\bf r} - {\bf r'})\delta(\omega - \omega')   \nonumber \\
 \equiv 2\pi\langle {{\bf j}^{\rm f}_i({\bf r})}{{{\bf j}}^{\rm f}_{k}({\bf r'})^*}\rangle_\omega \delta(\omega - \omega')  \label{1}
\end{align}
where the last equality defines the spectral function (the Fourier transform of $\langle {{\bf j}^{\rm f}_i({{\bf r}}, t)}{{\bf j}^{\rm f}_k({{\bf r'}}, t')}\rangle$ with respect to the time difference $(t - t'$).




 In Eq.~(\ref{1}) a local isotropic dielectric function was assumed. For the general case of a nonlocal anisotropic response $\varepsilon({\rm r, \omega})\delta_{ik}\delta({\rm r - \rm r'})$ should be replaced by  $\varepsilon({\rm r, \rm r', \omega})$.

 Eq.~(\ref{1}) makes it obvious that Rytov's method requires that some of the bodies under consideration (at least one) must be dissipative (i.e., have a nonzero imaginary part of the dielectric function), serving as sources for the fluctuating fields. For instance, in the Lifshitz problem one of the semi-infinite spaces (or both) must be dissipative- otherwise there are no sources in the Maxwell equations and the fluctuation force is manifestly zero. It is interesting, though, that in the final expression for the force, i.e., after Wick rotating to the imaginary $\omega$-axis, one can set the dielectric functions to be strictly real everywhere and still obtain a finite value for the force. It turns out, thus, that it is essential to keep in Rytov's treatment  at least an infinitesimal  dissipation which can be set to zero only at the end of the calculation. This point has been addressed in Ref.\cite{brevik21} and we will illustrate it in Sec.II by deriving the vacuum fluctuations within the Rytov's approach.
 %
 Curiously, for $T=0$, Lifshitz's result can also be derived using a quantum electrodynamics (QED) treatment ignoring dissipation effects, and supposing $\epsilon(\omega)$ is strictly real-valued for all real (positive) $\omega$.
 Even though the idealization of a lossless dispersive response may be useful in some contexts, realistic materials are dissipative. Moreover, a lossless dispersion is incompatible with the Kramers-Kronig relations (as the Kramers-Kronig theory assumes from the outset that there are no poles on the real frequency axis) as well as with some exact sum rules that any realistic  $\epsilon(\omega)$ must obey. Although this issue has been extensively discussed in \cite{brevik21}, we briefly re-iterate some points in Sec.III.

 A topic in the ``Casimir physics'', which is nowadays under active research, is the so called ``quantum friction''  \cite{dedkov,intra1,Milton,intra2}. Examples of this phenomenon are the friction force experienced by two plates in a shear motion or a nanoparticle moving parallel to a material plate.
 Interestingly, dissipation is typically an essential ingredient to ensure that the friction force is associated with a stationary process. We illustrate this point in Sect. IV, showing that for sufficiently low dissipation the friction force may exhibit an exponential growth in time.
 Instead of moving the particle relative to the plate, one can consider a setup, perhaps more realistic, when  the particle and the plate are both at rest but an electric current is flowing through the plate, thus exerting a tangential force on the particle (and, in addition, modifying the normal Casimir-Lifshitz force acting on the particle).  The Rytov's method is very suitable for studying this kind of problems. This will be done in Sec. V where some references will be provided.

   \section{The Quantum Vacuum as a Fluctuating Medium}

   Quantum vacuum ($T = 0$) is a ``medium'' with strictly real $\epsilon$ = 1 and, at the first sight, the fluctuational electrodynamics, with its inevitable necessity for a finite ${\mathop{\rm Im}\nolimits} \left\{ {\varepsilon \left( \omega  \right)} \right\}$ should be inapplicable to the vacuum. It is well known, however, that vacuum does exhibit a fluctuating electromagnetic field. It is perhaps less known that the properties of this field can be obtained within the fluctuational electrodynamics formalism, in the appropriate limit. We start with the more general problem of an arbitrary (but homogeneous and isotropic)  material medium \cite{landau60, landau80}. The dielectric function of the medium is a complex scalar  $\epsilon(\omega)$ independent of {\bf r} (we assume the medium to be non-magnetic). The fluctuating currents   in the medium produce an electric (and magnetic) field which (in the frequency domain) satisfies
 \begin{equation}
-\frac{\omega^2\epsilon(\omega)}{c^2}{\bf E({\bf r}, \omega)}  + \nabla  \times \nabla  \times {\bf E} =  \frac{4\pi i\omega}{c^2}{\bf j}^{\rm f}({\bf r}, \omega)   \label{2}
\end{equation}
Fourier transforming (2) in space  we obtain the following relation between
${\bf E({\bf k}, \omega)}$ and ${{\bf j}^{\rm f}({\bf k}, \omega)}$:
\begin{equation}
{\bf E({\bf k}, \omega)} = \frac{4\pi i}{\omega\epsilon(\omega)}\frac{\epsilon(\omega)k_0^2{{\bf j}^{\rm f}({\bf k}, \omega)}- ({\bf k}\cdot {{\bf j}^{\rm f}({\bf k}, \omega))}{\bf k}} {k^2 - k_0^2\epsilon(\omega) }   \label{3}
\end{equation}
where $k_0=\omega/c$.


Eq (3), with the use of (1) (transformed to ${\bf k}$-representation), enables one to compute the correlation function $\langle {{\bf E}_i({\bf k}, \omega)}{{\bf E}^*_k({\bf k'}, \omega')}\rangle$. Then, returning to real space, one finds the spectral density $\langle {{\bf E}_i({\bf r})}{{\bf E}^*_k({\bf r'})}\rangle_\omega$ \cite{landau60,landau80}. We only write the final expression for the field spectral density with contracted indices, i.e. (see Eq. (77.8) in \cite{landau80})

\begin{equation}
 \langle {\bf E({\bf r})}\cdot {\bf E({\bf r'})}\rangle_\omega = 2\hbar \coth \left( \frac{\hbar \omega}{2T}\right) {\rm{Im}}\left[ \frac{\omega^2}{c^2 R}\exp(-\frac{\omega R}{c}\sqrt {-\epsilon}) +       \frac{2\pi}{\epsilon}\delta (R)\right] \label{4}
\end{equation}
where $R = |{\bf r} - {\bf r'}|$.  Eq (4) is applicable to any absorbing medium and in its derivation it was necessary to have some finite ${\mathop{\rm Im}\nolimits} \left\{ {\varepsilon } \right\}$. Technically, this was necessary when calculating an integral over $k$ while  the integrand contains a denominator  ($k^2 - k_0^2\epsilon(\omega)$).  Note, however, that after the integral is performed, one can set in the final expression (4) $\epsilon$ identical to a real number (the limit of transparent medium). Thus, in Rytov's formalism, when treating a transparent medium, one still needs to keep an infinitesimal ${\rm Im} \left(\epsilon\right)$ which can be set to zero only at the end, after the thermodynamic limit is taken \cite{landau80}. In particular, if we take in (4) $T=0$ and $\epsilon=1$, we obtain the vacuum fluctuations
\begin{equation}
\langle {\bf E({\bf r})}\cdot {\bf E({\bf r'})}\rangle_\omega = \frac{2\hbar\omega^2}{c^2R} \sin(\frac{\omega R}{c}).   \label{5}
\end{equation}
There is an identical expression for the spectral density of the magnetic field in vacuum. These electromagnetic field fluctuations are responsible for the Casimir-Lifshitz forces in vacuum, at $T=0$ or, more realistically, for temperatures smaller than $\hbar\omega_0$ where, for a dielectric material $\omega_0$ corresponds to a frequency region where strong absorption happens, usually at optical frequencies (for metals $\omega_0$ corresponds to the plasma frequency).



Thus, in fluctuational electrodynamics vacuum is treated as a uniform isotropic medium in the limit ${\rm Im} \left(\epsilon\right)\rightarrow 0$, ${\rm Re}(\epsilon)=1$, $T=0$ (It is essential, though, not to set ${\rm Im} \left(\epsilon\right)=0$ from the beginning: this would totally suppress the fluctuating currents in Rytov's approach). It is, in fact, quite remarkable that in this way one can obtain the same result as in a quantum electrodynamics calculation, where one computes the vacuum expectation value for the corresponding operators (of course, there is no dissipation at all in that calculation but, still, an infinitesimal "decay" is introduced by selecting the retarded Green's function or by assigning an infinitesimal imaginary part to the frequency).

\section{Various Models of Materials}

The Lifshitz theory, based on Rytov's fluctuational electrodynamics, is considered the cornerstone for calculations of the Lifshitz-Casimir forces, for arbitrary materials and at arbitrary temperatures. It turns out, however, that in many cases theoretical calculations are in disagreement with experimental results  \cite{mostepanenko21}. The experiments are claimed to be ``precision experiments'' so, assuming this is indeed the case, the problem must be with the theory. In order to obtain theoretical results, which can be compared with experiments, one must, of course,  consider a specific model for the relevant material, with some definite dielectric function $\varepsilon(\omega)$ or, more generally, $\varepsilon(\omega, {\bf k})$, if the spatial dispersion is included. Thus, if the Lifshitz theory, based on the  very general and well established FDT, disagrees with the "precision experiments", the most reasonable explanation is that the employed model does not correctly describe the actual material.

All real materials are dissipative, to one degree or another.
Indeed, as clearly stated for instance in the textbook \cite{landau84}, p.280 ``...the imaginary part of $\varepsilon$ is positive for positive real omega  i.e. on the right-hand half of the real axis.''
Without dissipation a steady-state excitation with a bounded amplitude may result in a response with unbounded amplitude. This point can be nicely illustrated by a simple LC-circuit excited at resonance: the voltage and current amplitudes in the circuit grow linearly in time. In a real-system this is not possible due to the presence of nonlinearities (e.g., coupling with phonons), which lead to frequency conversion removing energy from the main harmonic preventing its growth beyond some threshold.

Sometimes neglecting dissipation, in certain frequency  intervals (far away from the absorption bands of the material) can be a very good approximation.
A textbook example is reflection and refraction in a transparent material.
Some authors employ in their studies the completely dissipation-less plasma model (DPM) \cite{mostepanenko21}. Ref.  \cite{brevik21} contains an extensive analysis of the deficiencies of the DPM. Below we briefly discuss some material models and introduce a few concepts and equations that will be needed later.

The DPM postulates a strictly real
\begin{equation}
	\varepsilon_p(\omega)= 1-\frac{\omega_p^2}{\omega^2},
	\label{6}
\end{equation}
where $\omega_p = (4\pi e^2 n_0/m)^{1/2}$ is the plasma frequency ($n_0$ is the electron concentration).  The expression (6) can be a good approximation for $\epsilon(\omega)$ in a collision-less plasma at high frequencies and with Landau damping being neglected. However, for real metals, it is unacceptable at low frequencies where spatial dispersion sets in and (6) must be replaced by a tensor $\varepsilon_{ik}(\omega, {\bf k})$ whose components depend on both $\omega$ and ${\bf k}$. The physical reason for such behavior, as clearly explained in \cite {lifshitzkinetics}, is that at low frequencies the wavelength becomes smaller than the oscillation amplitude of the electrons and the response becomes nonlocal. It follows, thus, that the second order pole in (\ref{6}) is an artifact
due to the use of a model  which is inadmissible at low frequencies.

A more realistic model, often considered on the par with the DPM \cite{mostepanenko21} is the Drude model, with
\begin{equation}
	\varepsilon_D(\omega)= \varepsilon_L(\omega)-\frac{\omega_p^2}{\omega[ \omega + i\gamma(T)]}, \label{7}
\end{equation}
which does allow for dissipation via the relaxation frequency $\gamma(T)$ which can depend on temperature $T$. The term $\varepsilon_L(\omega)$ accounts for the polarization of the lattice.

Both the DPM and the Drude model have been extensively studied in connection with the Casimir-Lifshitz forces. It has been argued that, somewhat surprisingly,  the less realistic DPM is in better agreement with experiment than the Drude model \cite{mostepanenko21}.
Moreover, the latter violates the Nernst heat theorem (if $\gamma(T)$ approaches zero faster than linearly) while the former is free from this deficiency \cite{footnote}.
Thus, as emphasized in  \cite{brevik21}, neither DPM nor the Drude model (for $\gamma(T)$ approaching zero with $T$) are  satisfactory and one must resort to models with spatial dispersion. In fact the importance of spatial dispersion has been recognized already in the early work on fluctuational electrodynamics. In recent years there have been some efforts to derive a realistic expression for the tensor $\varepsilon_{ik}(\omega, {\bf k})$ and to use it for calculating the Casimir-Lifshitz forces \cite{dalvit08,svetovoy,sernelius05,svetovoy05,pitaevskii08,davies}.

\section{Nonperturbative Quantum Friction}
A single body, moving with constant velocity in vacuum ($T=0$) does not experience any force (vacuum is Lorentz invariant!). However, two bodies moving in vacuum with a constant velocity relative to one another do experience a  friction force ("quantum friction").
For a long time studies of this effect yielded controversial results and some authors even denied the very existence of quantum friction \cite{leo}. In an important paper Pendry \cite{pen} (based on the earlier paper by the same author \cite{pen1})
 pointed out the error in argumentation of \cite{leo} and gave a clear intuitive picture of the effect, supported by a rigorous calculation based on perturbation theory. As mentioned above, at present the mechanism of this phenomenon is firmly established  \cite{intra1,Milton,dedkov,intra2}.
The two common geometries are a plate sliding on top of another plate (with no mechanical contact) or a nanoparticle (atom) moving parallel to a material plate.

The quantum friction effect has its origin in the spontaneous conversion of the kinetic energy of the moving body into light \cite{Silveirinha_Spontaneous}, which eventually is dissipated in the form of heat.
Typically, the quantum friction effect is a stationary process in the sense that the force is independent of time in time intervals where the variation of the velocity of the moving bodies can be ignored. Here, we point out that for weakly dissipative bodies it is possible to have a situation where the rate of emission of photons exceeds the rate of absorption so that the shear motion of two bodies leads to an electromagnetic instability. Such unstable regime has been discussed in a number of previous works both classically and with quantum methods \cite{Maslovski, Silveirinha1, Silveirinha2, Silveirinha3, Lannebere, Maghrebi}.
Here, building on \cite{Maslovski, pen} we consider a simple geometry that allows for a fully analytical (non-perturbative) quantum description of the friction process in such unstable regime. We find that in the unstable regime the friction force grows exponentially and leads to the emission of coherent light, analogous of a ``laser'' pumped by mechanical motion.
\begin{figure}[htbp]
	\centering
	 \includegraphics[width=0.5\linewidth,clip]{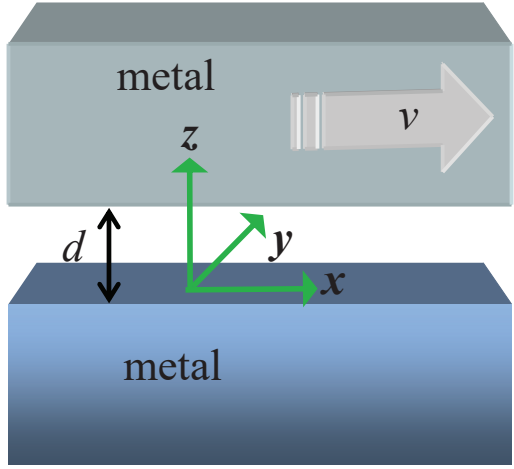}
	\caption{Two metal plates moves are sheared with velocity $v$. The distance between the plates is $d$.}
	\label{figgeom}
\end{figure}
We consider the quantum friction problem for two identical metal sheets separated by a distance $d$, analogous to problem discussed in Ref. \cite{pen1} (Fig. \ref{figgeom}). The relative velocity of the plates is ${\bf{v}} = v {{\bf{\hat x}}}$. In a quasi-static approximation, the metal sheets interact through the surface plasmons (SPPs). Following Pendry \cite{pen1}, in the quasi-static regime the system is described by the Hamiltonian:
\begin{equation}
\hat H = \sum\limits_{\bf{k}} {\hbar {\omega _{{\rm{sp}}}}\left( {{{\hat a}^\dag }_{{\bf{k}},1}{{\hat a}_{{\bf{k}},1}} + \frac{1}{2}} \right)}  + \sum\limits_{\bf{k}} {\hbar {\omega _{{\rm{sp}}}}\left( {{{\hat a}^\dag }_{{\bf{k}},2}{{\hat a}_{{\bf{k}},2}} + \frac{1}{2}} \right)}  + {\hat H_{{\mathop{\rm int}} }}.
\end{equation}
In the above, $\omega_{\rm {sp}}=\omega_p / \sqrt 2$ stands for the surface plasmon resonance, ${{{\hat a}_{{\bf{k}},l}}}$ and ${\hat a_{{\bf{k}},l}^\dag }$ with $l=1,2$ are creation and anihilation operators for plasmons with in-plane wave vector ${\bf{k}} = \left( {{k_x},{k_y},0} \right)$. The creation and annihilation operators satisfy the usual canonical commutation relations. Similar to Pendry, for simplicity, we consider the limit of vanishing material loss. However, it is underlined that the unstable regime persists when material dissipation is included in the calculation \cite{Silveirinha3}. Specifically, there is a threshold for the collision frequency in the metal below which the friction force exhibits the exponential growth discussed below.

The Hamiltonian that describes the interaction of the two plates is:
\begin{equation}
{\hat H_{{\mathop{\rm int}} }} = \sum\limits_{\bf{k}} {\frac{{\hbar {\omega _{{\rm{sp}}}}}}{2}\left( {{{\hat a}_{{\bf{k}},1}} + \hat a_{ - {\bf{k}},1}^\dag } \right)\left( {\hat a_{{\bf{k}},2}^\dag  + {{\hat a}_{ - {\bf{k}},2}}} \right)} {e^{ - {k_{||}}d}}{e^{ + i{\bf{k}} \cdot {\bf{v}}t}},
\end{equation}
with ${k_{||}} = \left| {\bf{k}} \right| = \sqrt {k_x^2 + k_y^2}$. The creation and annihilation operators are related to those in Ref. \cite{pen1}  $\hat a_{c{\bf{k}}}$ and $\hat a_{s{\bf{k}}}$, as follows  ${\hat a_{c{\bf{k}}}} = i\frac{{{{\hat a}_{\bf{k}}} + {{\hat a}_{ - {\bf{k}}}}}}{{\sqrt 2 }}$, ${\hat a_{s{\bf{k}}}} = \frac{{{{\hat a}_{\bf{k}}} - {{\hat a}_{ - {\bf{k}}}}}}{{\sqrt 2 }}$. A factor of 2 was suppressed in the formula of Ref. \cite{pen1} (it can be checked that with the additional factor of 2 the interaction of the slabs would lead to a detuning of the SPP resonance $\omega_{\rm {sp}}$ two times stronger than what is classically expected in the absence of relative motion).

In Ref. \cite{pen1} the friction force was found using Fermi’s golden rule to calculate the transitions from the ground state to excited states. Next, we solve the same problem \emph{exactly}, without using any perturbation formalism. The analysis extends to the quantum case the classical treatment of Sect. V of Ref. \cite{Maslovski}. To begin with, we note that the time dynamics of the annihilation operators is determined by $\frac{{d{{\hat a}_{{\bf{k}},l}}}}{{dt}} = \frac{i}{\hbar }\left[ {\hat H,{{\hat a}_{{\bf{k}},l}}} \right] =  - i{\omega _{{\rm{sp}}}}{\hat a_{{\bf{k}},l}} + \frac{i}{\hbar }\left[ {{{\hat H}_{{\mathop{\rm int}} }},{{\hat a}_{{\bf{k}},l}}} \right]$ so that:

\begin{equation}
i\frac{{d{{\hat a}_{{\bf{k}},1}}}}{{dt}} = {\omega _{{\rm{sp}}}}{\hat a_{{\bf{k}},1}} + \frac{{{\omega _{{\rm{sp}}}}}}{2}\left( {\hat a_{ - {\bf{k}},2}^\dag  + \hat a_{{\bf{k}},2}^{}} \right){e^{ - {k_{||}}d}}{e^{ - i{{\bf{k}}} \cdot {\bf{v}}t}}, \nonumber
\end{equation}
\begin{equation}
i\frac{{d{{\hat a}_{{\bf{k}},2}}}}{{dt}} = {\omega _{{\rm{sp}}}}{\hat a_{{\bf{k}},2}} + \frac{{{\omega _{{\rm{sp}}}}}}{2}\left( {\hat a_{ - {\bf{k}},1}^\dag  + \hat a_{{\bf{k}},1}^{}} \right){e^{ - {k_{||}}d}}{e^{ + i{{\bf{k}}} \cdot {\bf{v}}t}}. \nonumber
\end{equation}
Introducing $\hat b_{{\bf{k}},1}^{} = \hat a_{{\bf{k}},1}^{}{e^{ + i{{\bf{k}}} \cdot {\bf{v}}t/2}}$  and $\hat b_{{\bf{k}},2}^{} = \hat a_{{\bf{k}},2}^{}{e^{ - i{{\bf{k}}} \cdot {\bf{v}}t/2}}$, it is simple to check that the above equations are equivalent to the following system:
\begin{equation}
i\frac{d}{{dt}}\left( {\begin{array}{*{20}{c}}
{\hat b_{{\bf{k}},1}^{}}\\
{\hat b_{{\bf{k}},2}^{}}\\
{\hat b_{ - {\bf{k}},1}^\dag }\\
{\hat b_{ - {\bf{k}},2}^\dag }
\end{array}} \right) = \underbrace {\left( {\begin{array}{*{20}{c}}
{{\omega _{{\rm{sp}}}} - \frac{{{\bf{k}} \cdot {\bf{v}}}}{2}}&{\frac{{{\omega _{{\rm{sp}}}}}}{2}{e^{ - {k_{||}}d}}}&0&{\frac{{{\omega _{{\rm{sp}}}}}}{2}{e^{ - {k_{||}}d}}}\\
{\frac{{{\omega _{{\rm{sp}}}}}}{2}{e^{ - {k_{||}}d}}}&{{\omega _{{\rm{sp}}}} + \frac{{{\bf{k}} \cdot {\bf{v}}}}{2}}&{\frac{{{\omega _{{\rm{sp}}}}}}{2}{e^{ - {k_{||}}d}}}&0\\
0&{ - \frac{{{\omega _{{\rm{sp}}}}}}{2}{e^{ - {k_{||}}d}}}&{ - \left( {{\omega _{{\rm{sp}}}} + \frac{{{\bf{k}} \cdot {\bf{v}}}}{2}} \right)}&{ - \frac{{{\omega _{{\rm{sp}}}}}}{2}{e^{ - {k_{||}}d}}}\\
{ - \frac{{{\omega _{{\rm{sp}}}}}}{2}{e^{ - {k_{||}}d}}}&0&{ - \frac{{{\omega _{{\rm{sp}}}}}}{2}{e^{ - {k_{||}}d}}}&{ - \left( {{\omega _{{\rm{sp}}}} - \frac{{{\bf{k}} \cdot {\bf{v}}}}{2}} \right)}
\end{array}} \right)}_{\bf{M}}\left( {\begin{array}{*{20}{c}}
{\hat b_{{\bf{k}},1}^{}}\\
{\hat b_{{\bf{k}},2}^{}}\\
{\hat b_{ - {\bf{k}},1}^\dag }\\
{\hat b_{ - {\bf{k}},2}^\dag }
\end{array}} \right)
 \label{system}
\end{equation}
Since the (non-Hermitian) matrix on the right-hand side is independent of time, the time evolution of the operators can be found exactly. The behavior of the operators is  controlled by the four eigenvalues of the matrix:
\begin{equation}
{\omega _{\bf{k}}} =  \pm \sqrt {\omega _{{\rm{sp}}}^2 + {{\left( {\frac{{{{\bf{k}}} \cdot {\bf{v}}}}{2}} \right)}^2} \pm \omega _{{\rm{sp}}}^2\sqrt {{e^{ - 2{k_{||}}d}} + {{\left( {\frac{{{\bf{k}} \cdot {\bf{v}}}}{{\omega _{{\rm{sp}}}^{}}}} \right)}^2}} }. \nonumber
\end{equation}
The solution of the differential system (\ref{system}) is given by (the matrix ${S_{\bf{k}}}$ below  is formed by the eigenvectors of the matrix in $\bf{M}$ in Eq. (\ref{system}) and effectively determines a Bogoliubov transformation; note that the matrix $\bf{U}$ is given by ${\bf{U}}\left( t \right) = \exp \left( { - it{\bf{M}}} \right)$):
\begin{equation}
\left( {\begin{array}{*{20}{c}}
{\hat b_{{\bf{k}},1}^{}\left( t \right)}\\
{\hat b_{{\bf{k}},2}^{}\left( t \right)}\\
{\hat b_{ - {\bf{k}},1}^\dag \left( t \right)}\\
{\hat b_{ - {\bf{k}},2}^\dag \left( t \right)}
\end{array}} \right) = \underbrace {{S_{\bf{k}}}\left( {\begin{array}{*{20}{c}}
{{e^{ - i{\omega _{\bf{k}}}_ + t}}}&0&0&0\\
0&{{e^{ - i{\omega _{\bf{k}}}_ - t}}}&0&0\\
0&0&{{e^{ + i{\omega _{\bf{k}}}_ - t}}}&0\\
0&0&0&{{e^{ + i{\omega _{\bf{k}}}_ + t}}}
\end{array}} \right) \cdot S_{\bf{k}}^{ - 1}}_{\bf{U}} \cdot \left( {\begin{array}{*{20}{c}}
{\hat b_{{\bf{k}},1}^{}\left( 0 \right)}\\
{\hat b_{{\bf{k}},2}^{}\left( 0 \right)}\\
{\hat b_{ - {\bf{k}},1}^\dag \left( 0 \right)}\\
{\hat b_{ - {\bf{k}},2}^\dag \left( 0 \right)}
\end{array}} \right)
\end{equation}
In the above, ${\omega _{\bf{k}}}_ \pm $  represent the eigenvalues ${\omega _{{\bf{k}} \pm }} = \sqrt {\omega _{{\rm{sp}}}^2 + {{\left( {\frac{{{{\bf{k}}_{}} \cdot {\bf{v}}}}{2}} \right)}^2} \pm \omega _{{\rm{sp}}}^2\sqrt {{e^{ - 2{k_{||}}d}} + {{\left( {\frac{{{{\bf{k}}_{}} \cdot {\bf{v}}}}{{\omega _{{\rm{sp}}}^{}}}} \right)}^2}} }$.
There are two branches due to the hybridization of the plasmons in the two metal sheets.
For most wave vectors, ${\omega _{\bf{k}}}_ \pm $  are positive numbers. However, for certain wave vectors one of the branches (${\omega _{\bf{k}}}_ -$) may become complex-valued. In such a case, the matrix $\bf U$ exhibits exponential growth which indicates a parametric instability and the generation of quanta \cite{Maslovski}.

The range of wave vectors that leads to the instability satisfies (weak interaction is assumed):
\begin{equation}
2 - {e^{ - {k_{||}}d}} < \left| {\frac{{{{\bf{k}}_{}} \cdot {\bf{v}}}}{{\omega _{{\rm{sp}}}^{}}}} \right| < 2 + {e^{ - {k_{||}}d}}, \quad \rm{(unstable\,range\,of\,wave\, vectors).} \label{unstable}
\end{equation}
For wave vectors in the interval (\ref{unstable}), the corresponding eigenvalues satisfy:
\begin{equation}
{\omega _{{\bf{k}} - }} \approx  + i\frac{{\omega _{{\rm{sp}}}^{}}}{2}{e^{ - {k_{||}}d}}\sqrt {1 - {{\left( {\frac{{\left| {{k_x}} \right| - 2\frac{{{\omega _{{\rm{sp}}}}}}{{\left| v \right|}}}}{{\frac{{{\omega _{{\rm{sp}}}}}}{{\left| v \right|}}{e^{ - {k_{||}}d}}}}} \right)}^2}}  \equiv i{\omega ''_{{\bf{k}} - }}.
\label{wli2}
\end{equation}
As seen, in the unstable range the relevant eigenvalues are purely imaginary. It should be noted that the time variation of the original creation and annihilation operators ${{{\hat a}_{{\bf{k}},l}}}$ and ${\hat a_{{\bf{k}},l}^\dag }$ has an additional $\exp\left(\pm i k_x\, v\,t/2\right)$ factor which in the unstable range is approximately $\exp\left(\pm i \omega_{\rm sp} t\right)$. Due to this reason the emitted quanta have frequency $\omega_{sp}$ in the laboratory frame (where one of the plates is at rest), i.e., the instability is associated with the generation of plasmons.

\begin{figure}[htbp]
	\centering
	 \includegraphics[width=0.65\linewidth,clip]{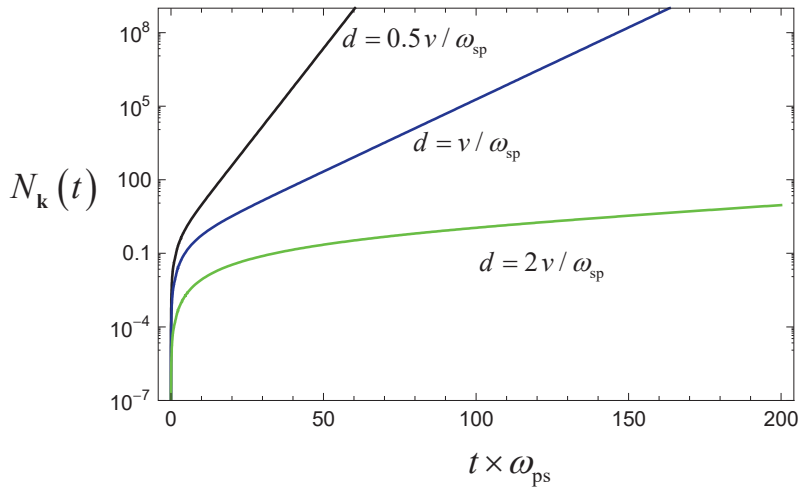}
	\caption{Number of quanta as a function of time for the physical channel with the largest growth rate ($k_y=0$, $k_x = 2 \omega_{\rm{sp}}/v$) and different distances between the metal plates. Notice that the vertical scale is in logarithmic units.}
	\label{figForce}
\end{figure}

The number of quanta in surface 1 associated with normal modes with wave vectors $\bf{k}$ and $-\bf{k}$ is given by (${U_{ij,{\bf{k}}}}\left( t \right)$ denotes the $ij$ element of the matrix $\bf U$):
\begin{equation}
\begin{array}{l}
{N_{{\bf{k}},1}}\left( t \right) = \left\langle {0|\hat b_{{\bf{k}},1}^\dag \left( t \right)\hat b_{{\bf{k}},1}^{}\left( t \right) + \hat b_{ - {\bf{k}},1}^\dag \left( t \right)\hat b_{ - {\bf{k}},1}^{}\left( t \right)|0} \right\rangle \\
{\rm{           }} = {\left| {{U_{13,{\bf{k}}}}\left( t \right)} \right|^2} + {\left| {{U_{14,{\bf{k}}}}\left( t \right)} \right|^2} + {\left| {{U_{13, - {\bf{k}}}}\left( t \right)} \right|^2} + {\left| {{U_{14, - {\bf{k}}}}\left( t \right)} \right|^2}
\end{array} \nonumber
\end{equation}
The number of quanta in surface 2 is defined in a similar way. It can be shown that ${N_{{\bf{k}},1}}\left( t \right) = {N_{{\bf{k}},2}}\left( t \right)$. The numerically calculated time variation of ${N_{{\bf{k}},1}}\left( t \right) = {N_{{\bf{k}},2}}\left( t \right) \equiv {N_{\bf{k}}}\left( t \right)$  is depicted in Fig. \ref{figForce} for the case $k_y=0$, $k_x = 2 \omega_{\rm{sp}}/v$, which corresponds to the physical channel that maximizes the growth rate. As seen, the number of quanta grows exponentially in time. This is due to the spontaneous conversion of kinetic energy into radiation which in this problem leads to coherent light emission \cite{Maslovski, Silveirinha1, Silveirinha2, Silveirinha3}.
In particular, the total energy (due to the quanta generated in the two slabs) grows in time as $E\left( t \right) = \sum\limits_{{k_x} > 0,{k_y}} {2\hbar {\omega _{{\rm{sp}}}}{N_{\bf{k}}}\left( t \right)}$  with the sum restricted to the unstable wave vector range with $k_x>0$.

The friction force per unit of area is given by (below $A$ represents the area of the plates):
\begin{equation}
\frac{{{F_{{\rm{fr}}}}}}{A} = \frac{1}{A}\sum\limits_{{k_x} > 0,{k_y}} {\frac{{2\hbar {\omega _{{\rm{sp}}}}}}{{\left| v \right|}}\frac{{d{N_{\bf{k}}}\left( t \right)}}{{dt}}}. \label{friction}
\end{equation}
Clearly, similar to the number of quanta, the friction force grows exponentially with time. The exponential growth of the force leads to an exponential reduction of the relative velocity of the two bodies, which weakens the instability. Furthermore, material nonlinearities may also lead to a saturation effect that stops the instability.

It is interesting to compare the ``exact'' formula (\ref{friction}) with the result obtained perturbatively with the Fermi's golden rule. Specifically, it is shown next that Pendry's theory can be recovered using  $\frac{{d{N_{\bf{k}}}\left( t \right)}}{{dt}} \approx \frac{1}{{{\tau _{\bf{k}}}}}$, where $\tau _{\bf{k}}$ is the inverse of the grow rate of the relevant natural mode: ${\tau _{\bf{k}}} = \frac{1}{{2{{\omega}_{{\bf{k}} - }''}}}$. Note that $\tau _{\bf{k}}$ is roughly the time required to generate a single quanta of light with wave vector $\bf k$ from the quantum vacuum state. This approximation leads to a stationary (time-independent) force:
\begin{equation}
\frac{{{F_{{\rm{fr}}}}}}{A} \approx \frac{1}{A}\sum\limits_{{k_x} > 0,{k_y}} {\frac{{2\hbar {\omega _{{\rm{sp}}}}}}{{\left| v \right|}}2{{\omega}_{{\bf{k}} - }''} = } \frac{{2\hbar {\omega _{{\rm{sp}}}}}}{{\left| v \right|}}\frac{1}{{{{\left( {2\pi } \right)}^2}}}\int\limits_{{k_x} > 0} {\int\limits_{ - \infty }^{ + \infty } {d{k_x}d{k_y}} } \,\, 2{\omega ''_{{\bf{k}} - }}
\end{equation}
Using Eq. (\ref{wli2}), it is possible to perform the integral in $k_x$ analytically (see also a related calculation in Sect. V of Ref. \cite{Maslovski}). This yields:
\begin{equation}
\frac{{{F_{{\rm{fr}}}}}}{A} \approx \frac{{\hbar \omega _{{\rm{sp}}}^3}}{{4\pi {v^2}}}\int\limits_{ - \infty }^{ + \infty } {d{k_y}} {e^{ - 2\sqrt {k_y^2 + {{\left( {2{\omega _{{\rm{sp}}}}/v} \right)}^2}} d}}.
\end{equation}
This is precisely the result of Pendry derived with perturbation theory \cite{pen1} (the result of Pendry is two times larger because his interaction Hamiltonian is also two times larger). Clearly, the perturbation approach does not capture correctly the time dynamics of the friction force in the unstable regime. A comparison between Pendry's result at $T=0$ and other approaches can be found in Ref. \cite{Milton}. Moreover, the study of the friction force can be extended to finite temperatures, see for example \cite{Hoye}. It should be noted that for the model considered here, the friction force falls off exponentially with $v$ rather than the usual $v^3$ law \cite{Hoye}. Such a feature is likely a consequence of (i) the quasi-static approximation which assumes that the friction force arises only due to the interaction of plasmons, and neglects other (low-frequency) physical channels for the interaction, and (ii) the fact that the unstable regime considered here is different from the ``stationary'' regime considered in other works where the friction force is independent of time.

In summary, the interaction of two moving bodies may lead to a quantum friction force for vanishingly small material dissipation. The effect is rooted in a parametric instability that leads to the coherent emission of light. Different from most of the models in the literature, our theory highlights that the friction force does not need to be associated with a stationary process, but may rather exhibit an exponential growth before some nonlinear process kicks in to stop the instability.

\section{Non-contact Friction in various setups}


In this section, we concentrate on a less studied setup, when the bodies are at rest but an electrical current is flowing in one of them.

For a finite temperature $T$,  the inter-body space is not a vacuum but is filled by thermal radiation so that, in addition to "quantum friction" the bodies experience also "thermal friction"\cite{pod}. Moreover, the temperatures of different bodies need not be the same. Obviously, one is dealing here with a non-equilibrium situations of moving bodies, with different temperatures, so that the use of Eq.~(\ref{1}) needs some justification. The requirement is that each body separately must be in its internal equilibrium, with a well defined temperature.

\begin{figure}[htbp]
	\centering
	 \includegraphics[width=0.6\linewidth,clip]{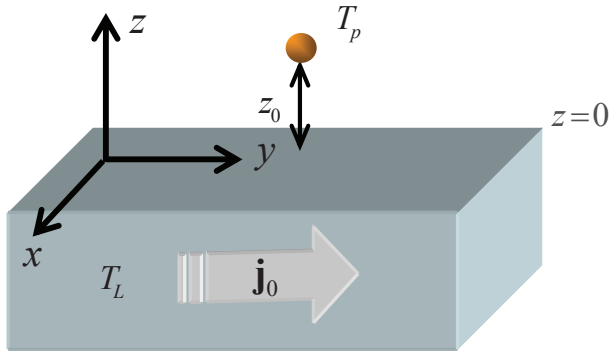}
	\caption{A small particle is situated at a distance $z_0$ from the surface of a plate. Both the particle and the plate are at rest but a dc current is flowing through the plate. The temperature of the plate, $T_L$, is generally different from that of the particle, $T_p$.}
	\label{figDrift}
\end{figure}

Specifically we consider a small particle (it can be an atom, a molecule or a nanoparticle) in close proximity to a conducting plate (Fig. \ref{figDrift}). Both the particle and the plate are at rest (in the lab reference frame) . The plate is electrically biased so that there is  a dc current with density ${\bf j}_{0}=en_{0}{\bf v}_{0}$  in the sample. Here ${\bf v}_{0}$ is the drift velocity of the carriers, $e$ is their charge and
$n_{0}$ is the average concentration. The fluctuating currents ${{\bf j}^{\rm f}({\bf r}, t)}$ and fields ${{\bf E}({\bf r}, t)}$ now are "riding" on top of this stationary drift current which affects the correlation function  of the fields, both inside and outside the sample. This fact has been pointed out long ago \cite{s1,s3} but it is only more recently that the effect of the carrier drift on the fluctuation-induced  forces came into study \cite{vol1,vol2,s2}

The carrier drift modifies the usual (i.e. equilibrium) force but also produces a lateral drag force. Below we briefly describe how this happens (for details see \cite{s2}).

First, the dielectric function, which relates the fluctuating parts of the electric displacement and the electric field, now becomes a tensor

\begin{equation}
	\epsilon_{ik}\left(\omega,{\bf k}\right)=\epsilon{}_{L}
	\delta_{ik}-\frac{\omega_{p}^{2}}{\omega\left(\omega-{\bf k}\cdot{\bf v}_{0}+i\gamma\right)}\left(\delta_{ik}+
	\frac{v_{0i}k_{k}}{\omega-{\bf k}\cdot{\bf v}_{0}}\right).\label{epsilon}
\end{equation}

This is the extension of the Drude model to the case when the carrier drift is present. The latter is responsible for the non-local relation between the fluctuating field and current (spatial dispersion).

Second, the correlation function of the spontaneous fluctuating currents must be modified. More precisely there are two sources of fluctuations: (i) The lattice has some losses, i.e. some imaginary part ${\rm{Im}}\,\varepsilon_L(\omega)$ in its dielectric constant. The corresponding fluctuations are not affected by the drift (the lattice is at rest, at some temperature $T_L$)
(ii) The electronic component is noisy due to losses described by the decay rate $\gamma$. However, since the fluctuating currents now originate in  a frame moving with respect to the lab frame, one should Doppler shift the corresponding correlation function, i.e. replace $\omega$ by $\omega-{\bf k}\cdot{\bf v}_{0}$. Moreover, the electron component might have its own temperature, $T_{el}$, different from that of the lattice.

Third, for a particle in close proximity to the surface  fluctuations are dominated by  the near field (evanescent waves). In this case one can deal with the Poisson equation, instead of the full set of Maxwell equations, and only the longitudinal component of  $\epsilon_{ik}\left(\omega,{\bf k}\right)$)
\begin{equation}
	\epsilon\left(\omega,{\bf k}\right)=
	\epsilon_{L}\left(\omega\right)-\frac{\omega_{p}^{2}}
	{\left(\omega_{-}+i
		\gamma\right)\left(\omega_{-}\right)}\label{epsilonl}
\end{equation}
appears in the calculation. Here $\omega_{-} = \omega - {\bf k}\cdot{\bf v}_{0}$ is the Doppler shifted frequency (compare to Eq.(\ref{7})).

The electric response of the particle  is described by its susceptibility
\begin{equation}
	\alpha\left(\omega\right) =
	\frac{\alpha\left(0\right)\omega_{0}^{2}}{\omega_{0}^{2}-\omega^{2}-i\omega\eta}.\label{alpha}
\end{equation}
For an atom, modeled as a two-level system, the resonant frequency $\omega_{0}$ is the energy spacing between the levels. For a metallic or semiconducting (spherical)
nanoparticle,  $\alpha\left(0\right)$ is equal to the
cube of the radius of the sphere and
$\omega_{0}=\tilde\omega_{p}/\sqrt{3}$ is the frequency of the
localized surface plasmon \cite{mai}. (Here $\tilde\omega_{p}$ is
the plasma frequency of the material of the particle). For a
dielectric nanoparticle one may have some phonon mode or a
phonon-polariton, instead of a plasmon. Finally, $\eta$ designates the decay rate of the excitation.

Let us stress that in our setup only the electronic component of the plate (the electron plasma) is moving with respect to the particle while the lattice (as the particle itself) is stationary in the lab frame. Therefore our results differ from those obtained for the "usual" setup (the entire plate is dragged with respect to the particle). Moreover, our results strongly depend on whether the main source of fluctuations resides in the lattice or in the electron plasma. The relation between the three temperatures,  $T_L$, $T_{el}$ and the particle temperature $T_p$ is also an important factor. Below, as an illustration, we give the final result for the drag force $F_x$ for the case when the fluctuations originate in the electron plasma while fluctuations in the lattice are neglected (model 2 in \cite{s2}).

\begin{equation}
	\begin{array}{c}
		F_{x}\left(z_{0}\right)=\frac{\hbar}{\pi^{2}}\int_{0}^{\infty}d\omega\alpha^{\prime\prime}\left(\omega\right)
		\int\int_{-\infty}^{\infty}dk_{x}dk_{y}
		\left[\coth\left(\frac{\hbar\omega_{-}}{2T_{el}}\right) -
		\coth\left(\frac{\hbar\omega}{2T_{p}}\right)\right]\\
		\rm{Im}\Gamma\left(\omega,
		k_{x}\right)qk_{x}e^{-2qz_{0}}.\label{Fx}
	\end{array}
\end{equation}
where $z_0$ is the distance of the particle from the plate surface and

\begin{equation}
	\Gamma\left(\omega, k_{x}\right)=\frac{\epsilon\left(\omega, k_{x}\right)-1}{\epsilon\left(\omega, k_{x}\right)+1}.\label{G}
\end{equation}

We will not pursue the analysis of Eq.(\ref{Fx}), except for the following comments:

(i) The existence of the drag force requires that both the plate and the particle must be dissipative.

(ii) Since the electron plasma is drifting, the frequency in the argument of the corresponding $\coth$ is $\omega_{-}$ (Doppler shift) while the second $\coth$, corresponding to the lattice, contains just $\omega$.

(iii) Eq.(\ref{Fx}), unlike its counterpart for the case when fluctuations are dominated by the lattice, does resemble the result for the ``usual'' setup (if the fluctuations are dominated by the electron plasma). Indeed, the only essential difference between the two setups is that in our case the (passive) lattice is stationary with respect to the particle.

(iv) In the limit $T_p =T_{el} =0$ (quantum drag) the analysis simplifies and the important parameter ($v_0/\omega_{0}z_0)\equiv \kappa$ emerges in a clear way. For small values of the parameter (small velocities) $F_x\sim v_0^3$ while for large velocities , $v>>\omega_{0}z_0$,  the drag force drops as $1/v^2$.

As has been already mentioned, the drift current (in addition to producing the drag force) also gives corrections to the normal component, $F_z$ (the Casimir-Lifshitz force). We will not dwell on this issue.

\section{conclusion}

We have discussed various aspects of the theory of fluctuation induced forces, including examples and some novel results. On the conceptual side, we have emphasized the subtlety of the dissipation-less limit in the fluctuational electrodynamics. Namely, if one wants to approach the limit of a lossless transparent medium (vacuum included), one must first introduce infinitesimal losses, assigning some small imaginary part $\rm{Im}\epsilon$ to the dielectric function, and only at the end  (after the thermodynamic limit is taken) one can set  $\rm{Im}\epsilon$ to zero. Thus, the thermodynamic limit and the $\rm{Im}\epsilon\rightarrow 0$ do not commute. Disregard for this subtlety prompted some authors to introduce models with strictly real $\epsilon(\omega)$, like the dissipation-less plasma model \cite{mostepanenko21}. This model, in addition of having an nonphysical double pole at $\omega =0$, violates some exact relations and identities that any  material must obey.

The second part of the paper deals with the out-of-equilibrium phenomenon of "quantum friction" but goes beyond the usual situation, as described in the reviews \cite{dedkov,intra1,Milton,intra2}.   In particular, we point out that material dissipation may be an essential ingredient to have a stationary friction force. In a weakly dissipative system, the friction force may be associated with an electromagnetic instability such that the kinetic energy, related to the relative motion of two bodies,
 is transformed into an exponentially growing coherent radiation. This is a subtle nonperturbative  effect whose complete theory is given in Sec.IV. Finally, we have briefly addressed the problem of  the fluctuation-induced drag force, acting on a small polarizable
neutral particle (atom, molecule or a nanoparticle), situated in the vicinity of a body through which a constant electrical current is flowing. Here (unlike the more common setup when the particle and the body are in relative motion) both objects are at rest and only the electronic component (the electron plasma) in the body is moving. Thus, although in some cases the results for the forces  in the two setups can be similar, generally they are quite different.

\begin{acknowledgements}
M.S. was partially funded by  the Institution of Engineering and Technology (IET) under the A F Harvey Research Prize 2018 and by Instituto de Telecomunicações under Project Number UID/EEA/50008/2020.
\end{acknowledgements}


\begin{thebibliography}{99}
\bibitem{footnote} In practice, even in the $T\rightarrow 0$ limit  the Drude model contains some residual $\gamma$, due to static impurities, which rectifies the problem with the Nernst theorem \cite{hoye03,brevik06})
\end{thebibliography}
\end{document}